\documentclass[conference]{IEEEtran}

\usepackage{amsmath}
\usepackage{color}
\usepackage{listings}
\usepackage{amsfonts}
\usepackage{graphicx}
\usepackage[hidelinks]{hyperref}
\usepackage[utf8]{inputenc}
\usepackage[hang,small]{caption}

\begin{document}
%
\title{Multi-core computation of transfer matrices for strip lattices in the Potts model}

\author{\IEEEauthorblockN{Cristobal A. Navarro}
\IEEEauthorblockA{Department of Computer Science\\
Universidad de Chile\\
Santiago, Chile\\
Email: crinavar@dcc.uchile.cl}
\and
\IEEEauthorblockN{Fabrizio Canfora}
\IEEEauthorblockA{Centro de Estudios científicos (CECS)\\
Valdivia, Chile\\
Email: canfora@cecs.cl}
\and
\IEEEauthorblockN{Nancy Hitschfeld}
\IEEEauthorblockA{Department of Computer Science\\
Universidad de Chile\\
Santiago, Chile\\
Email: nancy@dcc.uchile.cl}}

\maketitle
\begin{abstract}
The transfer-matrix technique is a convenient way for studying strip lattices in the Potts model since the computational costs depend just on the periodic part of the lattice 
and not on the whole. 
However, even when the cost is reduced, the transfer-matrix technique is still an \textit{NP-hard} problem since the time $T(|V|,|E|)$ needed to compute the matrix grows exponentially as a function of the graph width. 
In this work, we present a parallel transfer-matrix implementation that scales performance under multi-core architectures. 
The construction of the matrix is based on several repetitions of the deletion-contraction technique, allowing parallelism suitable to multi-core machines. 
Our experimental results show that the multi-core implementation achieves speedups of $3.7X$ with $p=4$ processors and 
$5.7X$ with $p=8$. The efficiency of the implementation lies between 60\% and 95\%, achieving the best balance of speedup and efficiency at $p=4$ processors for actual multi-core architectures. The algorithm also takes advantage of the lattice symmetry, making the transfer matrix computation to run up to $2X$ faster than its non-symmetric counterpart and use up to a quarter of the original space.
\end{abstract}
\IEEEpeerreviewmaketitle

\section{Introduction}
The Potts model \cite{potts} has been widely used to study physical phenomena of \textit{spin lattices} such as 
phase transitions in the thermo-dynamical equilibrium. Topologies such as 
triangular, honeycomb, square, kagome among others are of high interest and are being studied frequently
(see \cite{Chang_Shrock_2000, Shrock_Tsai_1999, Chang_Salas_Shrock_2002, Chang_Jacobsen_Salas_Shrock_2002}). When the number of 
possible spin states is set to $q=2$, the Potts model becomes the classic Ising model \cite{ising_1925} which has 
been solved analytically for the whole plane by Onsager \cite{NYAS:NYAS627}. 
Unfortunately, for higher values of $q$ no full-plane solution has been found yet. Therefore, studying strip lattices becomes
a natural approach for achieving an exact but finite representation of the bidimensional plane. The wider the strip, 
the better the representation. Hopefully, by increasing the width enough, 
some properties of the full plane would emerge. 

One known technique for obtaining the partition function of a strip lattice 
is to compute a transfer matrix based on the periodic information of the system. One should be aware however that building the transfer matrix is not free 
of combinatorial computations and exponential cost algorithms. In fact, the problem requires the computation 
of partition functions which are \textit{NP-hard} problems \cite{Woeginger:2003:EAN:885909.885927}.

With the evolution of multi-core CPUs towards a higher amount of cores, parallel computing is not anymore limited to clusters or super-computing; workstations can also 
provide high performance computing. It is in this last category where most of the scientific community lies, therefore parallel implementations for multi-core machines are the ones to have 
the biggest impact. Latest work in the field of the Potts model has been focused on parallel probabilistic simulations 
\cite{ferrero2011, DBLP:journals/cphysics/TapiaD11} and new sequential methods for computing the exact partition function 
of a lattice \cite{hartmann-2005-94, bedini, Shrock_2000}. To the best of our knowledge, there has not been published research regarding parallel multi-core performance of exact transfer-matrix algorithms in the Potts model. The closest related research regarding this subject has been the massive parallel GPU implementations of the Monte Carlo algorithms 
\cite{ferrero2011, 4724234}, which is out of the scope of this research since they are not exact.
Even when the exact methods have much higher cost than probabilistic ones, they are still important because 
one can obtain exact behavior of the thermo-dynamical properties of the system such as the free energy, magnetization and specific heat. 
Once the matrix is computed, it can be evaluated and operated as many times as needed. It is important then to provide a fast way for computing the matrix in its symbolic form and not numerically, since the latter would imply a whole re-computation of the transfer matrix each time a parameter is modified. In this work, we have achieved an implementation that computes the symbolic transfer matrix and scales performance as more processors are available. The implementation can also solve problems larger than the system's available memory since it uses a secondary memory strategy, never storing the full matrix in memory.

The paper is organized as follows: section (\ref{seq_preliminaries_related_work}) covers preliminary concepts 
as well as related work, sections (\ref{seq_algorithm_overview}) and (\ref{seq_algorithm_optimizations}) explain 
the details of the algorithm and the additional optimizations to the implementation. 
In section (\ref{sec_performance}) we present experimental results such as run time, speedup, efficiency and knee, using different amount of processors. Section (\ref{seq_conclusions}) 
discusses our main results and concludes the impact of the work for practical usage.
\section{Preliminaries and related work}
\label{seq_preliminaries_related_work}
Let $G=(V,E)$ be a lattice with $|V|$ vertices, $|E|$ edges and $s_i$ be the state of a \textit{spin} of $G$ with 
$i \in [1..q]$. The Potts partition function $Z(G, q, \beta)$ is defined as
\begin{equation}
Z(G, q, \beta) = \sum_{r}e^{-{\beta}h(G_{r})}
\label{eq_potts}
\end{equation}
where $\beta = \frac{1}{K_B T}$, $K_B$ is the Boltzmann constant and $h(G_r)$ is the energy of the lattice at a given state 
$G_r$\footnote{A state $G_r$ is a distribution of spin values on the lattice. It can be seen as a graph with an specific 
combination of values on the vertices.}.
The Potts model defines the energy of a state $G_r$ with the following Hamiltonian:
\begin{equation}
\label{eq_hamiltonian}
h(G_r) = -J\sum_{\langle i,j\rangle \in G_r}\delta_{s_{i}, s_{j}}
\end{equation}
Where $\langle i,j\rangle$ corresponds to the edge from vertex $v_i$ to $v_j$, $r \in [1..q^{|V|}]$, $J$ is the interaction energy ($J<0$ for 
\textit{anti-ferromagnetic} and $J>0$ for \textit{ferromagnetic}) and $\delta_{s_{i}, s_{j}}$ corresponds to the \textit{Kronecker delta} evaluated at the pair of spins $\langle i, j \rangle$ with states $s_i,s_j$ and expressed as
\begin{equation}
\delta_{s_{i},s_{j}} = 
\left\{ 
\begin{array}{rl}
 1 &\mbox{ if $s_{i}=s_{j}$} \\
 0 &\mbox{ if $s_{i}\not=s_{j}$}
\end{array} 
\right.
\end{equation}
The free energy of the system is computed as:
\begin{equation}
F = -Tlog_e(Z)
\end{equation}
As the lattice becomes bigger in the number of vertices and edges, the computation of equation (\ref{eq_potts})
becomes rapidly intractable with an exponential cost of $\Theta(q^{|V|})$. In practice, an equivalent recursive method is more 
convenient than the original definition.

The \textit{deletion-contraction} method, or DC method, was initially used to compute the Tutte polynomial \cite{tutte} and 
was then extended to the Potts model 
after the relation found between the two (see \cite{potts_tutte_relation, sokal_2005}). 
DC re-defines $Z(..)$ as the following recursive equation
\begin{equation}
Z(G, q, v) = Z(G-e, q, v) + vZ(G/e, q, v)
\label{eq_deletion_contraction}
\end{equation}
$G-e$ is the \textit{deletion} operation, $G/e$ is the \textit{contraction} operation and 
the auxiliary variable $v = e^{-\beta J} - 1$ makes $Z(..)$ a polynomial.
There are three special cases when DC can perform 
a recursive step with linear cost:
\begin{equation}
Z(G, q, v)=
\left\{ 
\begin{array}{ll}
(q+v)Z(G/e, q, v) ;		&\mbox{if \{e\} is a spike.}\\
(1+v)Z(G-e, q, v) ;		&\mbox{if \{e\} is a loop.}\\
q^{|V|};			&\mbox{if $E=\{\emptyset\}$.}
\end{array} 
\right.
\label{eq_deletion_contraction_optimizations}
\end{equation}
The computational complexity of DC has a direct upper bound of $O(2^{|E|})$.
When $|E| >> |V|$ a tighter bound is known based on the 
Fibonacci sequence complexity \cite{Wilf:2002:AC:560438}; $O((\frac{1+\sqrt{5}}{2})^{|V|+|E|})$. 
In general, the time complexity of DC can be written as
\begin{equation}
T(G) = min\Bigg( O(2^{|E|}), O\Big(\frac{1+\sqrt{5}}{2}\Big)^{|V|+|E|}\Bigg)
\end{equation}
Haggard's et al. \cite{haggard_computing_tutte_polynomials} work is considered the best implementation of DC for computing 
the Tutte polynomial for any given graph.
Their algorithm, even when it is exponential in time, reduces the computation tree in the presence of loops, multi-edges, 
cycles and biconnected graphs (as one-step reductions). An important contribution by the authors is that by using a cache, 
some computations can be reused (i.e sub-graphs that are isomorphic to the ones stored in the cache do not need to be computed again).
An alternative algorithm was proposed by Bj{\"o}rklund et al.\cite{DBLP:journals/corr/abs-0711-2585} 
which accomplishes exponential time only on the number of vertices;
$O(2^nn^{O(1)})$ with $n=|V|$. Asymptotically their method is better than DC
considering that many interesting lattices have more edges than vertices. 
However, Haggard \textit{et. al.} \cite{haggard_computing_tutte_polynomials} have stated 
that the memory usage of Bj{\"o}rklund's method is too high for practical usage.
For the case of strip lattices, a full application of DC is not practicable since the exponential cost would grow as a function of the total amount of edges, making the computation rapidly intractable. The \textit{transfer matrix} technique, 
mixed with DC is a better choice since the exponential cost will not depend on the total length of the strip, 
but instead just on the size of the period. 

As soon as the matrix is built, the remaining computations become numerical and less expensive; \textit{i.e.}, matrix multiplications (for finite length) or eigenvalue computations (for infinite length).
Bedini \textit{et. al.} \cite{bedini} proposed a method for computing the partition function of arbitrary graphs using a tree-decomposed transfer matrix. In their work, the authors obtain a sub-exponential algorithm based on arbitrary heuristics for finding a good tree decomposition of the graph. This method is the best known so far for arbitrary graphs, but when applied to strip lattices, it costs just as the traditional methods, \textit{i.e.}, the tree-width becomes the width of the strip and the cost is proportional to the Catalan number of the width. Therefore, it is of interest to use parallelism in order to improve the performance of the transfer matrix problem when dealing with strip lattices.

\begin{figure}[ht!]
\includegraphics[scale=0.4]{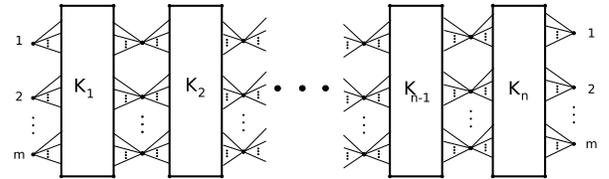}
\centering
\caption{Strip lattice model with length $n$ and width $m$.}
\label{fig_strip_lattice_general}
\end{figure}

A strip lattice is by default a bidimensional graph $G=(V,E)$ that is periodic at least along one dimension. 
It can be perceived as a concatenation of $n$ subgraphs $K$ sharing their boundary vertices and edges. 
Let $P$ be the set of all possible strip lattices, then we formally define $G$:
\begin{equation}
G\{V,E\} \in P\ \Leftrightarrow \ \exists K=\{V',E'\}: G=\bigotimes_{1}^nK
\label{eq_lattice_periodic_logic}
\end{equation}
$\bigotimes$ is the special operator for concatenating the periods defined as $K_i=(E_i,V_i)$ of height $m$. Each period 
connects to the other periods $K_{i-1}$ and $K_{i+1}$, except for $K_1$ and $K_n$ which are the external ones and connect to $K_2$ and 
$K_{n-1}$ respectively (see Figure \ref{fig_strip_lattice_general}).

The main computational challenge when using a matrix transfer based algorithm is the cost of building it 
because its size increases exponentially as a function of the width of the strip. The problem of the matrix size 
has been improved by analytic techniques \cite{MGhaemi}. However, the authors specify that these techniques are only applicable to 
square and triangular lattices using values of $q=2$ and $q=3$ (Ising and three-state Potts respectively). For this reason, we prefer to use a more general transfer matrix based on its combinatorial aspects, with the advantage 
of being useful to any lattice topology, and allowing any value of $q \in \mathbb{R}$.

To the best of our knowledge, there is no mention on the effectiveness on parallelizing transfer matrix algorithms for strip lattices in the Potts model. The core of our work is to focus on the multi-core parallel capabilities of a practical 
transfer matrix method and confirm or deny the factibility of such computation to run in parallel.
In order to achieve parallelism, we use a transfer matrix algorithm based on a modified \textit{deletion-contraction} (DC) scheme. It is in fact a \textit{partial DC} that stops its recursion when the edges to be processed connect a pair of vertices of the next period. As a result, the partial DC generates many partial partition functions associated to combinatorial labels located at the leaves of the recursion tree. These partial partition functions, when grouped by their combinatorial label, make a row of the transfer matrix. 
A hash table is a good choice for searching and grouping terms in the combinatorial space of the problem. 
\section{Algorithm overview}
\label{seq_algorithm_overview}
\subsection{Data structure}
Our definition of $K$ from section (\ref{seq_preliminaries_related_work}) (Figure \ref{fig_strip_lattice_general}) 
will be used to model our input data structure. Given any strip lattice $G$, 
only the right-most part of the strip lattice is needed, that is, $K_n$. 
We will refer to the data structure of $K_n$ as $K_{\sigma_1}$ to denote the basic case where the structure is equal to the original $K_n$, not having any additional modification.
For simplicity and consistency, we will use a \textit{top-down} enumeration of the vertices, such that the left 
boundary contains the first $m$ vertices and the right boundary the last $m$ ones. We will introduce the following naming 
scheme for left and right boundary vertices, this will be \textit{shared} and \textit{external} vertices, respectively. 
\textit{Shared vertices} correspond to the left-most ones, \textit{i.e.} the vertices that are shared with $K_{n-1}$ (see Figure \ref{fig_data_structure}) and 
are indexed top-down from $0$ to $m-1$.
\textit{External vertices} are the ones of the right side (\textit{i.e.}, the right end of the whole strip lattice) of $K_{\sigma_1}$ and are indexed top-down from $|V|-m-1$ to $|V|-1$. 
\begin{center}
	\includegraphics[scale=0.33]{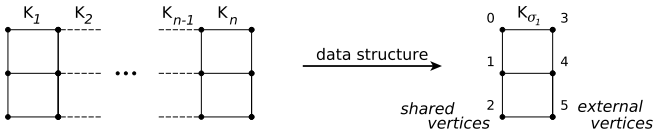}
	\captionof{figure}{Example data structure for a square lattice of width $m=3$.}
	\label{fig_data_structure}
\end{center}
\subsection{Computing the transfer matrix $M$.}
Computing the transfer matrix $M$ is a repetitive process that involves combinatorial operations over $K_{\sigma_1}$. 
For a better explanation of the algorithm, we introduce two terminologies; \textit{initial configurations} and 
\textit{terminal configurations}. These configurations 
define a combinatorial sequence of identifications for external and shared vertices, respectively, and correspond to the set of all non-crossing partitions. 
Given the lattice width $m$, the number of initial and terminal configurations is the sequence of the Catalan numbers:
\begin{equation}
C_m=\frac{1}{m+1}\binom{2m}{m}=\frac{(2m)!}{(m+1)!m!}=\prod_{k=2}^m\frac{m+k}{k}
\end{equation}
\textit{Initial configurations}, denoted $\sigma_i$ with $i \in [0..C_m-1]$, define a combinatorial sequence of identifications just on the external vertices. The 
\textit{terminal configurations}, denoted $\varphi_j$ with $j \in [0..C_m-1]$, define a combinatorial sequence of identifications just on the shared vertices. Initial 
configurations generate terminal ones (using the DC method) but not vice versa.
As stated before, $K_{\sigma_1}$ is the basic case and matches $K_n$. That is, $\sigma_1$ 
is the initial configuration where no identifications are applied to the external vertices. 
It is equivalent as saying that $\sigma_1$ is the empty partition of the Catalan set. Similarly, $\varphi_1$ corresponds to the base case where no shared vertices are identified. In other words, $\varphi_1$ is the empty configuration for the Catalan set on the shared vertices. 
Additionally, initial and terminal configurations have a maximum of $m$ vertices, and eventually will contain less vertices 
as more identifications are performed. 

The idea of the algorithm is to compute the transfer matrix $M$ in rows, by repeatedly applying 
the partial DC, each time to a different initial configuration $K_{\sigma_i}$, a total of $C_m$ times. Each repetition contributes to a row of $M$.
By default, the algorithm cannot know the $C_m$ different sequences of \textit{terminal} and \textit{initial} configurations except for $K_{\sigma_1}$ which is given as part of the input of the strip lattice and is the one that triggers the computation. This is indeed a problem 
for parallelization. 
To solve it, we use a recursive generator $g(A[\ ][\ ], s, H,S)$ that, with the help of a hash table $H$, generates 
all the $C_m$ configurations and stores them in an array $S$. $A[\ ][\ ]$ is an auxiliary array that stores the intermediate auxiliary subsequences and $s$ is the accumulated sequence of identifications. 
Before the first call of $g(A[\ ][\ ], s, H,S)$, $A = [[0,1,2,...,m-1]]$, $s$ is null and $H$ as well as $S$ are empty. 
$g(A[\ ][\ ], s, H,S)$ is defined as:
\begin{lstlisting}[frame=single]
g(A[][],s,H,S){
    if(!add_sequence(s,H,S))
        return
    for(int k=0; k<A.size(); k++){
    for(int j=1; j<A[k].size(); j++){
    for(int i=0; i<j; i++){
        if(can_identify(A[k],i,j)){
            cA := copy(A)
            cs := copy(s)
            identify(cA,i,j,k,cs)
            divide(cA,i,j,k)
            g(cA,cs,H,S)
        }
    }}}
}
\end{lstlisting}
Basically, $g(..)$ performs a three-way recursive division of the domain $A$. For each identification pair $i,j$, the domain is partitioned into three sets; (1) the top vertices above $i$, (2) the middle vertices between $i,j$ and (3) the vertices 
below $j$. If $|j-i| < 3$ then no set can be created in the middle. The same applies to the top and bottom sets if the distance from $i$ or $j$ to the boundary of the actual domain is less than 1. Each time a new identification $i,j$ is added, the resulting configuration is checked in the hash table. If it is a new one, then it is added, otherwise it is discarded as well as further recursion computations starting from that configuration. Thanks to the hash table, repetitive recursion branches are never computed.
Once $g(..)$ has finished, $S$ becomes the array of all possible configurations and $H$ is the hash that maps configurations 
to indices. 
At this point one is ready to start computing the matrix in parallel. We start dividing the total amount of rows by the amount of processors. Each processor 
will be computing a total of $C_m/p$ rows. The initial configuration sequence needed by each processor $p_i$ is obtained in parallel by reading from $S[p_i]$. Once the configuration is read, 
it is applied to the external vertices of its own local copy of the base case $K_{\sigma_1}$. 
After each processor builds their corresponding $K_{\sigma_i'}$ graph, each one performs a DC procedure in parallel, without any communication cost. This DC procedure is only partial 
because edges that connect two shared vertices must never be deleted neither contracted, otherwise one would be processing vertices and edges of the next period of the lattice, breaking the idea of a transfer matrix. An example of a partial DC is illustrated in Figure \ref{fig_example_recdc} for the case when computing the first row.
\begin{center}
	\includegraphics[scale=0.25]{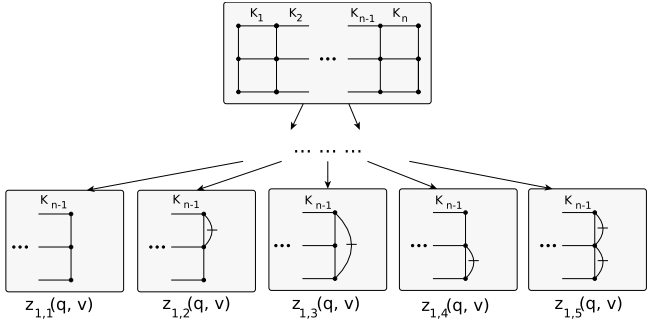}
	\captionof{figure}{An example of how terminal configurations are generated from the basic one.}
	\label{fig_example_recdc}
\end{center}
When the DC procedure ends, there will be partial expressions associated to a remanent of the graph at each leaf of the recursion tree. Each remanent corresponds to the part of the graph that was not computed (\textit{i.e}, edges connecting \textit{shared vertices}) and it is identified by its terminal configuration. Each one of these remanents specifies one of the $C_m$ possible terminal configurations that can exist. For some problems, not all \textit{terminal configurations} are generated from a single DC, but only a subset of them. That is why the generator function is so much needed in order for the algorithm to work in parallel, otherwise there would be a time dependency among the DC repetitions.

For each \textit{terminal configuration} $\varphi_j$, its key sequence is computed dynamically along the branch taken on the DC recursion tree; each contraction contributes with a pair of indices from the vertices. Consistent terminal configurations sequences are achieved by using a small algebra 
that combines the identifications from contractions, made on the \textit{shared vertices}. 
An identification of two shared vertices $[v_i,v_j]$ will be denoted as $\pi_{i,j}$. 
Each additional identification adds up to the previous ones to finally form a sequence of a terminal configuration
$\pi_{x_1,y_2} + \pi_{x_2, y_2} + ... + \pi_{x_n,y_n}$. The following properties hold true for sequences:
\begin{align}
\pi_{a,b} & = \pi_{b,a}
\label{eq_algebra_0}\\
\pi_{a,b}+\pi_{c,d} & = \pi_{c,d} + \pi_{a,b}
\label{eq_algebra_1}\\
\pi_{a,b}+\pi_{b,c} & = \pi_{a,b,c}
\label{eq_algebra_2}
\end{align}
If we apply a lexicographical order to each sequence,  we can avoid checking properties (\ref{eq_algebra_0}) and (\ref{eq_algebra_1}).

Each processor must group its partial expressions that share a common \textit{terminal configuration}, so that in the end there is only one final expression $z_{i,j}(q,v)$ per \textit{terminal configuration}. The list of final expressions associated to \textit{terminal configurations} represents one row of the transfer matrix. The final expressions become the elements of the row and the \textit{terminal configurations} are the keys for getting their respective column indices.
The terminal configuration sequence is necessary, but not sufficient for knowing its index $j$ in $M$. This is where $H$ becomes useful for knowing with average $O(1)$ cost what is the actual index $j$ of a given terminal configuration sequence. As a result, each thread $t_i$ can write their final expressions $z_{i,j}(q,v)$ correctly into $M$.
The main idea of the parallelization scheme can be illustrated by using Foster's \cite{fo_95} four-step strategy 
for building parallel algorithms; \textit{partitioning, communication, agglomeration, mapping}. Figure \ref{fig_foster_strategy_mp} 
shows an example using $p=2$.
\begin{center}
	\includegraphics[scale=0.2]{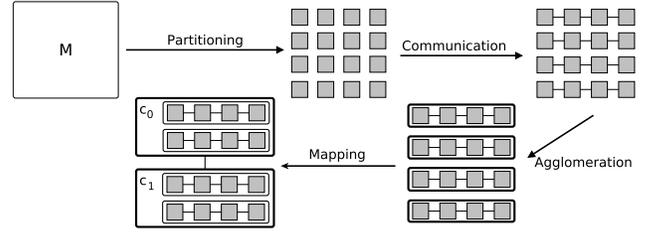}
	\captionof{figure}{The parallelism scheme under Foster's four step design strategy using two cores.}
	\label{fig_foster_strategy_mp}
\end{center}
Basically, the idea is to give a small amount of $B$ consecutive rows to each processor (for example $B=2,4,8 $ or $16$) with an offset 
of $k = pB$ rows per processor. If the work per row is unbalanced, 
then processing is better to be asynchronous, handling the work by a master process.
The asymptotic complexity for computing $M$ under the PRAM model using the CREW variation is upper-bounded by:
\begin{equation}
T(\vec{Z}) = O(\frac{C_m}{p}(min(2^{|E'|}, 1.6182^{|V'|+|E'|}))
\label{eq_transfer_matrix_complexity}
\end{equation}
The complexity equals the cost of applying DC times $C_m$ in parallel with $p$ processors.\\
When all processors end, the final transfer matrix $M$ is size $C_m$ x $C_m$:
\begin{equation}
M = 
\begin{vmatrix}
z_{1, 1}(q, v)	&	z_{1, 2}(q, v)	&	...	&	z_{1, Cm}(q, v)\\
z_{2, 1}(q, v)	&	z_{2, 2}(q, v)	&	...	&	z_{2, Cm}(q, v)\\
	...			&		...			&	...	&		...		     \\	
z_{Cm, 1}(q, v)	&	z_{Cm, 2}(q, v)	&	...	&	z_{Cm, Cm}(q, v)\\
\end{vmatrix}
\end{equation}
If the strip lattice represents an infinite band, then the next step is to make a numerical evaluation on $q,v$ and study the 
eigenvalues of $M$. If the strip lattice is finite, then a initial condition vector $\vec{Z_1}$ is needed. In that case, $M$ and $\vec{Z_1}$ together form a 
partition function vector $\vec{Z}$ based on the following recursion:
\begin{equation}
	\vec{Z}(n)=M\vec{Z}(n-1)
	\label{eq_Z_first}
\end{equation}
By solving the recurrence, $\vec{Z}$ becomes:
\begin{equation}
	\vec{Z} = M^{n-1}\vec{Z_1}
	\label{eq_Z}
\end{equation}
$\vec{Z_1}$ is computed by applying DC to each one of the $C_m$ \textit{terminal configurations}:
\begin{equation}
\vec{Z_1} = (DC(K_{\varphi_1}), DC(K_{\varphi_2}), ..., DC(K_{\varphi_{C_m}}))
\end{equation}
The computation of $\vec{Z_1}$ has very little impact on the overall cost of the algorithm. In fact the cost is practically $O(m)$
because a terminal configuration contains mostly \textit{spikes} and/or \textit{loops}, which are linear in cost. 
Moreover, \textit{terminal states} can be computed even faster by using the serial and parallel optimizations, but it is often not required. Computing the powers of $M^{n-1}$ should be done in a 
numerical context, otherwise memory usage will become intractable. 

Finally, the first element of $\vec{Z}$ is the partition function of studied the strip lattice.
After this point, $\vec{Z}$ is used to study a wide range of physical phenomena. Álvarez and Canfora \textit{et. al.} \cite{Alvarez_Canfora_Reyes_Riquelme_2012} 
have reported new exact results for strip lattices such as the kagome of width $m=5$ using the sequential version of this transfer matrix algorithm.
\section{Algorithm improvements}
\label{seq_algorithm_optimizations}
\subsection{Serial and Parallel paths}
The DC contraction procedure can be further optimized for graphs that present serial or parallel paths along their computation (see 
Figure \ref{fig_opt_sp}).
\begin{center}
	\includegraphics[scale=0.58]{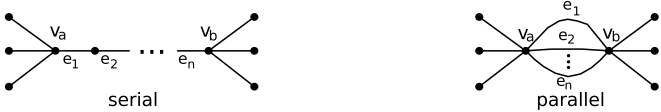}
	\captionof{figure}{Serial and parallel paths.}
	\label{fig_opt_sp}
\end{center}
Let $v_a$ and $v_b$ be the first and last vertices of a path, respectively. A \textbf{\textit{serial path} $s$} is a set of 
edges $e_1,e_2,...,e_n$ that connect sequentially $n-1$ vertices between $v_a$ and $v_b$.
It is possible to process a serial path of $n$ edges in one recursion step by using the following expression;
\begin{equation}
Z(K,q,v)=\Bigg[\frac{(q+v)^n - v^n}{q}\Bigg]Z(K_{-s},q,v) + v^nZ(K_{/s},q,v)
\end{equation}
\label{sec_optimizations}
On the other hand, a \textbf{\textit{parallel path} $p$} is a set of edges $e_1,e_2, ..., e_n$ where each one connects redundantly 
$v_a$ and $v_b$, forming $n$ possible paths between $v_a$ and $v_b$.
It is also possible to process a parallel path of $n$ edges in one recursion step by using the following expression;
\begin{equation}
Z(K,q,v)=Z(K_{-p},q,v) + \big[(1+v)^n - 1\big]Z(K_{/p},q,v)
\end{equation}
\subsection{Lattice Symmetry}
A very important optimization is to detect the lattice's symmetry when building the matrix. By detecting symmetry, the matrix size is 
significantly lower because all symmetric pair of \textit{terminal states} are grouped as one terminal state. As the width of the strip 
lattice increases, the number of symmetric states increases too, leading to matrices almost a half the dimension of the original $M$. 
We establish symmetry between two \textit{terminal configurations} $\varphi_a$ and $\varphi_b$ with 
keys $\pi_{a_1,...,a_n}$ and $\pi_{b_1,...,b_n}$ respectively in the following way:
\begin{equation}
\pi_{a_1,...,a_n} = \pi_{b_1,...,b_n} \Leftrightarrow a_i=(m-1)-b_{n-i+1} 
\end{equation}
Under symmetry, the resulting matrix has a different numerical sequence of sizes than the original $M$ which obeyed the Calatan numbers. 
In this case, we denote the size of the matrix as $D(m)$ and it is equal to:
\begin{align}
D(m) 	& = \frac{C_m}{2} + \frac{m!}{2\lfloor\frac{m}{2}\rfloor!}
\end{align}
As $m$ grows, the $\frac{C_m}{2}$ term increases faster than the second term. For big values of $m$, $D(m) \approx \frac{C_m}{2}$. 
Table (\ref{table_nonsym_sym_growth}) shows the rate at which a non-symmetric and symmetric matrix grows as $m$ increases.
\begin{table}[ht!]
\begin{center}
\caption{Growth rate of the size of $M$ under non-symmetric and symmetric cases.}
  \begin{tabular}{ c | c | c }
	m	& non-sym	& sym \\ 
	\hline
	2	& $2$		& $2$\\
	3 	& $5$ 	& $4$\\
   	4 	& $14$	& $10$\\
	5 	& $42$	& $26$\\
	6	& $132$	& $76$\\
	7	& $429$	& $232$\\
	8	& $1430$	& $750$\\
	9	& $4862$	& $2494$\\
	10	& $16796$	& $8524$
  \end{tabular}
  \label{table_nonsym_sym_growth}
\end{center}
\end{table}

\section{Implementation}
\label{sec_implementation}
We made two implementations of the parallel algorithm. One using OpenMP \cite{Chapman:2007:UOP:1370966} and the other one using MPI \cite{mpi}. We observed that the MPI implementation achieved 
better performance than the OpenMP one and scaled better as the number of processors increased. For this, we decided to continue the research with the MPI implementation and have discarded the OpenMP one. 
We chose a value of $B=4$ for the block-size (the amount of consecutive rows per process). The value was obtained experimentally by testing different values as powers of $2$, in the range $1 - 64$. As long as the parallelization 
is balanced a value of $B>1$ is beneficial.
An important aspect of our implementation is that we make each process generate its own $H$ table and $S$ array. This small sacrifice in memory leads to better performance than if $H$ and $M$ were shared 
among all processes. There are mainly three reasons why the replication approach is better than the sharing approach: (1) caches will not have to deal with consistency of shared data, (2) the cost of communicating the data structures is saved and (3) the allocation of the replicated data will be correctly placed on memory when working under a NUMA architecture. The last claim is true because on NUMA systems memory allocations on a given process are automatically placed in its fastest location according to the processor of the CPU. It is responsibility of the OS (or make manual mapping) to stick the process to the same processor through the entire computation.

The implementation saves each row to secondary memory as soon as it is computed. Each processor does this with its own file, therefore the matrix is fragmented into $p$ files. This secondary memory strategy is not a problem because practical case shows that numerical evaluation is needed before using the matrix in its non-fragmented form. In fact, fragmented files allow parallel evaluation of the matrix easier. We will not cover numerical evaluation in our experiments because it is out of the scope of this work.

\section{Experimental results}
\label{sec_performance}
We performed experimental tests for the parallel transfer matrix method implemented with MPI. 
The computer used for all tests is listed in table \ref{tab:testingPLatform}.
\begin{table}[ht]
{
\caption{Hardware and tools used for experiments.}
	\begin{center}
		\begin{tabular}{|l|l|}
			\hline
			Hardware & Detail \\
			\hline
			CPU & AMD FX-8350 8-core 4.0GHz \\
			Mem & 8GB RAM DDR3 1333Mhz\\
			MPI implementation & open-mpi\\
			\hline
		\end{tabular}
	\end{center}
\label{tab:testingPLatform}
}
\end{table}

Our experimental design consists of measuring the parallel performance of the implementation at 
computing the transfer matrix of two types of strip lattices (see Figure \ref{fig_test_lattices}); 
(1) \textit{square} and (2) \textit{kagome}.

\begin{center}
\includegraphics[scale=0.40]{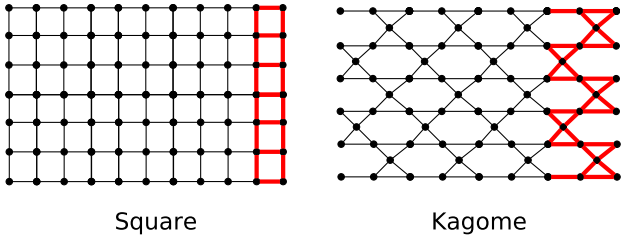}
\captionof{figure}{The two tests to be used for the experiments. The red part is the input for the program.}
\label{fig_test_lattices}
\end{center}

\subsection{Results on the square test}
For the test of the square lattice, we test 9 different strip widths in the range $m \in [2,10]$. 
For each width, we measure 8 average execution times, one for each value of $p \in [1,8]$. As a whole, 
we perform a total of 72 measurements for the square test. The standard error for the average execution times is below 5\%.
We also compute other performance measures such as speedup, efficiency and the \textit{knee} \cite{DBLP:journals/tc/EagerZL89}. In this tests, we made use of the lattice symmetry for all sizes of $m$.

\begin{figure*}[ht!]
\centering
\begin{tabular}{cc}
\includegraphics[scale=0.7]{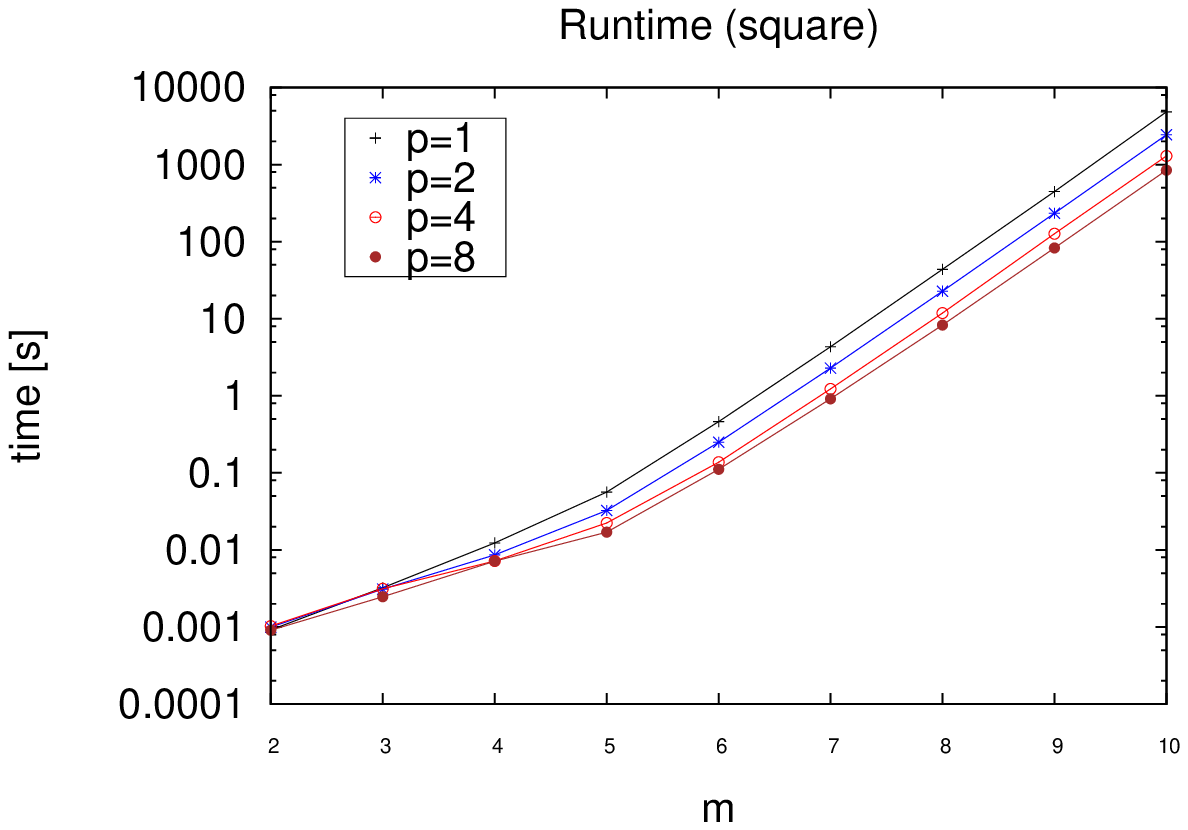} &
\includegraphics[scale=0.7]{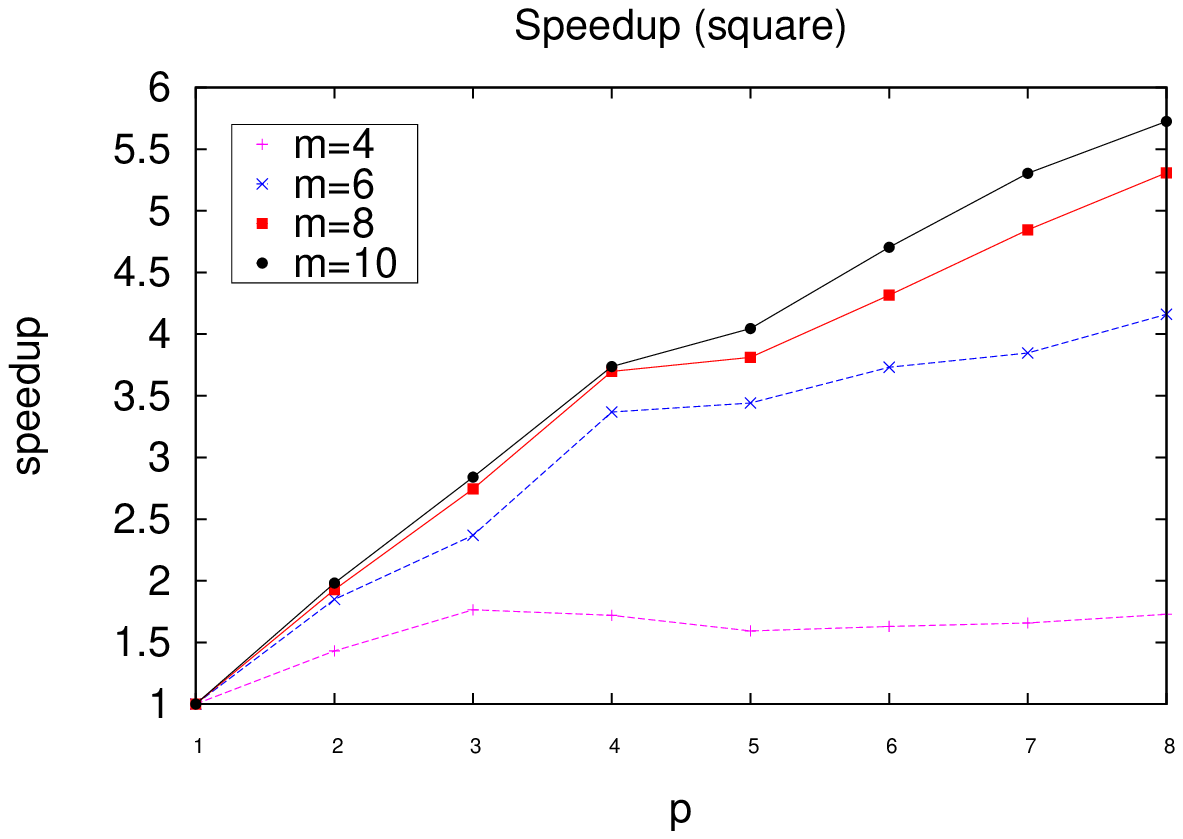}\\
\includegraphics[scale=0.7]{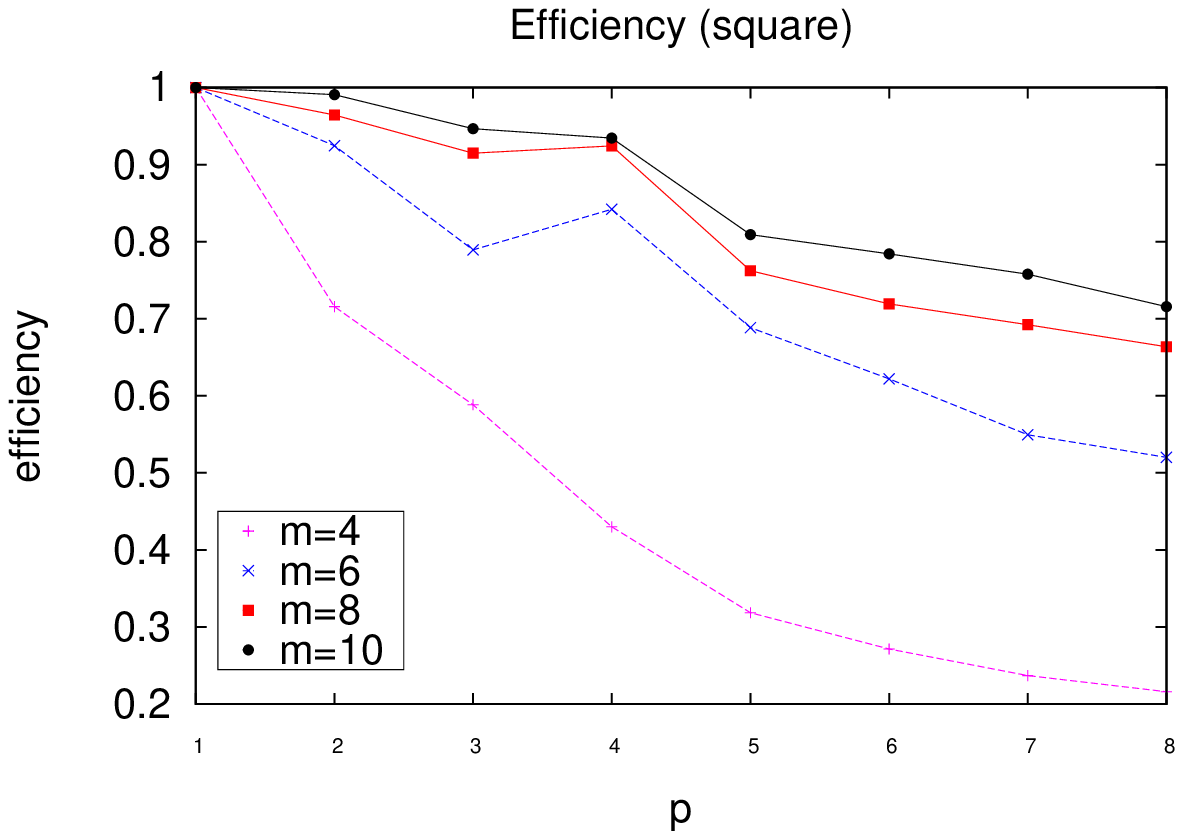} &
\includegraphics[scale=0.7]{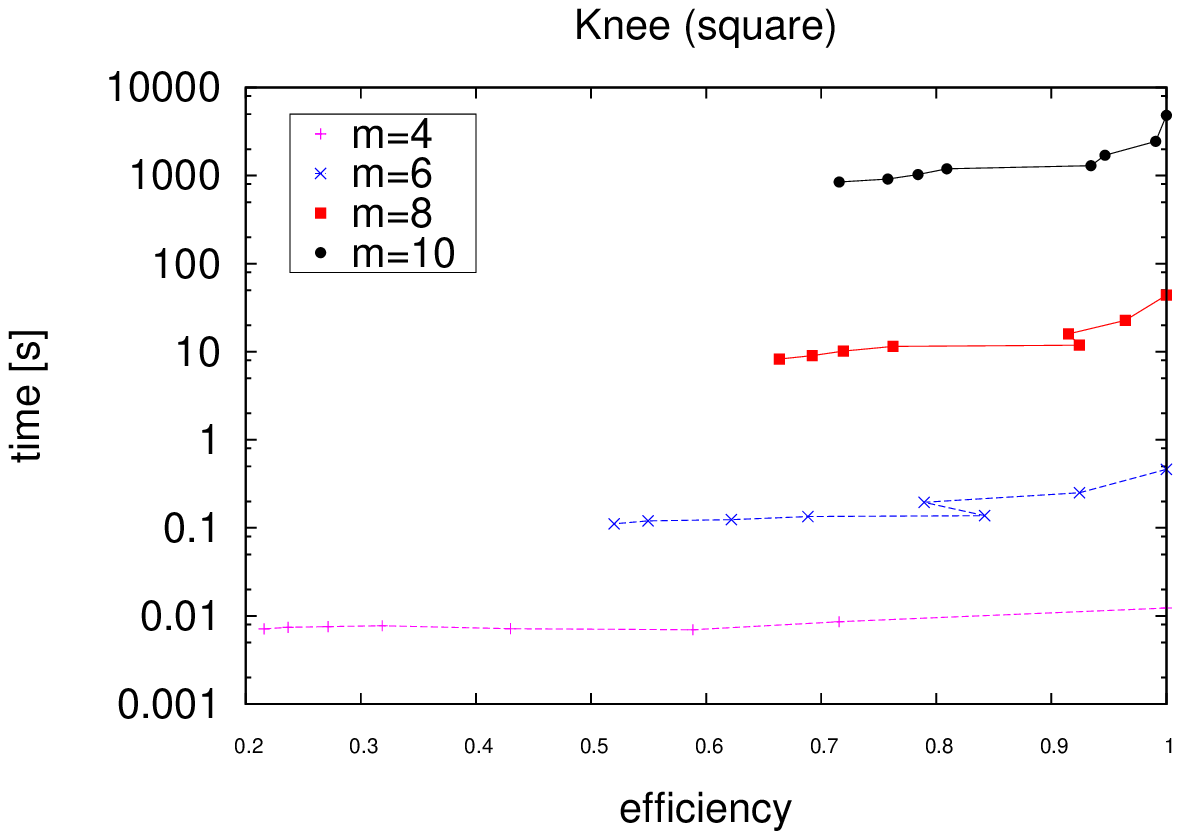}\\
\end{tabular}
\caption{Runtime, speedup, efficiency and knee for different sizes of the square strip lattice.}
\label{fig_square_performance_results}
\end{figure*}

Figure \ref{fig_square_performance_results} shows all four performance measures for the square lattice. 
From the results, we observe that there is speedup for every value of $p$ as long as $m > 4$. For $m \le 4$, the problem is 
not large enough to justify parallel computation, hence the overhead from MPI makes the implementation perform worse than 
the sequential version. The plot of the execution times confirms this behavior since the curves cross each other for $m < 4$.
The maximum speedup obtained was $5.7$ when using $p=8$ processors. From the lower left graphic we can see that efficiency 
decreases as $p$ increases, which is expected in every parallel implementation. What is important is that for large enough problems (\textit{i.e.}, $m>6$), efficiency is over 65\% for all $p$. For the case of $p=4$, we report 94\% of efficiency, which is close to perfect linear speedup. For $m \le 6$, the implementation is not so efficient because the amount of computation involved is not enough to keep all cores working at full capacity. 
The \textit{knee} is useful for finding the optimal value of $p$ for a balance between efficiency and computing time. 
It is called knee because the hint for the optimal value of $p$ is located in the knee of the curve (thought as a leg), that is, its lower right part. In order to know the value of $p$ suggested by the knee, one has to count the position of the closest point to the knee region, in reverse order. Our results of the knee for $m>6$ show that the best balance of performance and efficiency is achieved with $p=4$ (for $m \le 6$, the knee is not effective since there was no speedup in the 
first place). In other words, while $p=8$ is faster, it is not as efficient as with $p=4$.

\subsection{Results on the kagome test}
For the test of the kagome lattice, we used 5 different strip widths in the range $m \in [2,6]$. 
For each width, we measured 8 average execution times, one for each value of $p \in [1,8]$. As a whole, 
we performed a total of 40 measurements for the kagome test. The standard error for the average execution times is below 5\%.
Additional performance measures such as speedup, efficiency and knee have also been computed. In this 
test, we found that the block-size of $B=4$ was a bad choice because the work per row was unbalanced. Instead, we found by experimentation that $B=1$ makes the work assignation much more balanced. In this test we can only use lattice symmetry for 
$m= 3, 5$. We decided to run the whole benchmark without symmetry in order to maintain a consistent behavior for all values of $m$ (but we will still report how it performs when using symmetry).

\begin{figure*}[ht!]
\centering
\begin{tabular}{cc}
\includegraphics[scale=0.7]{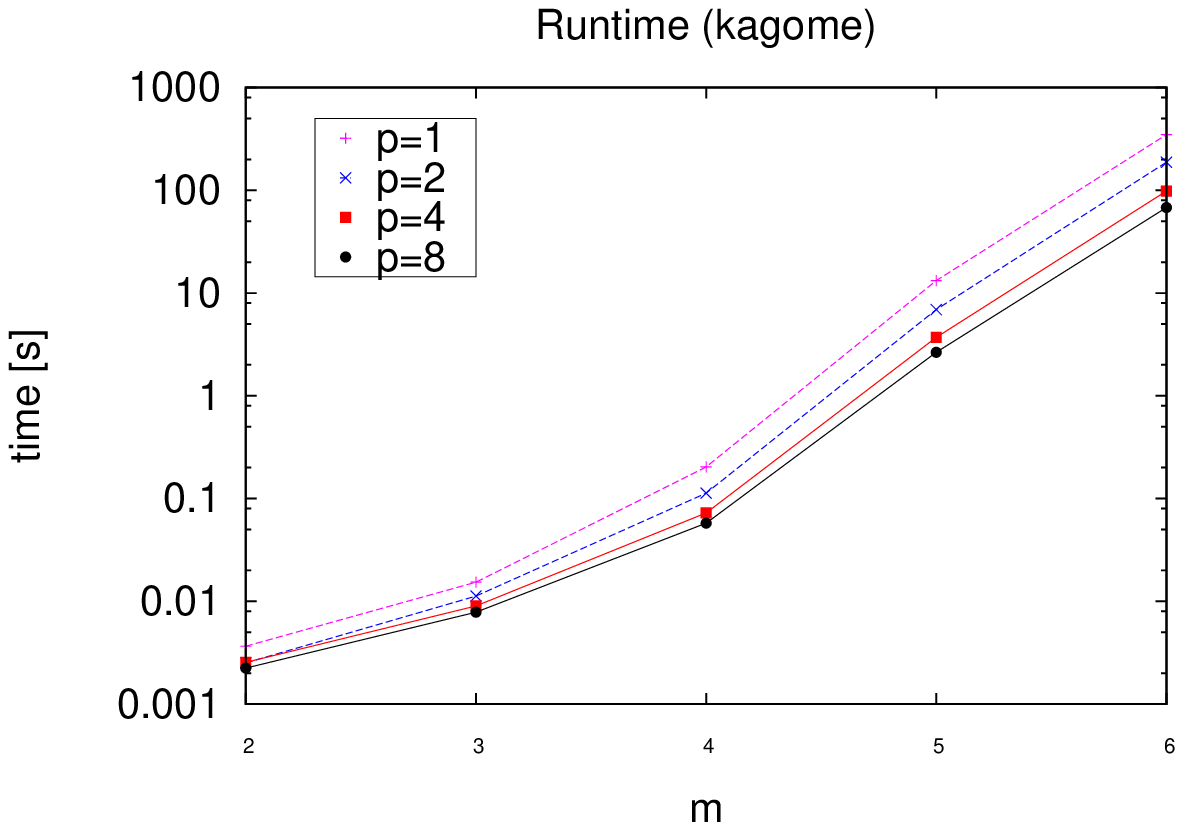} &
\includegraphics[scale=0.7]{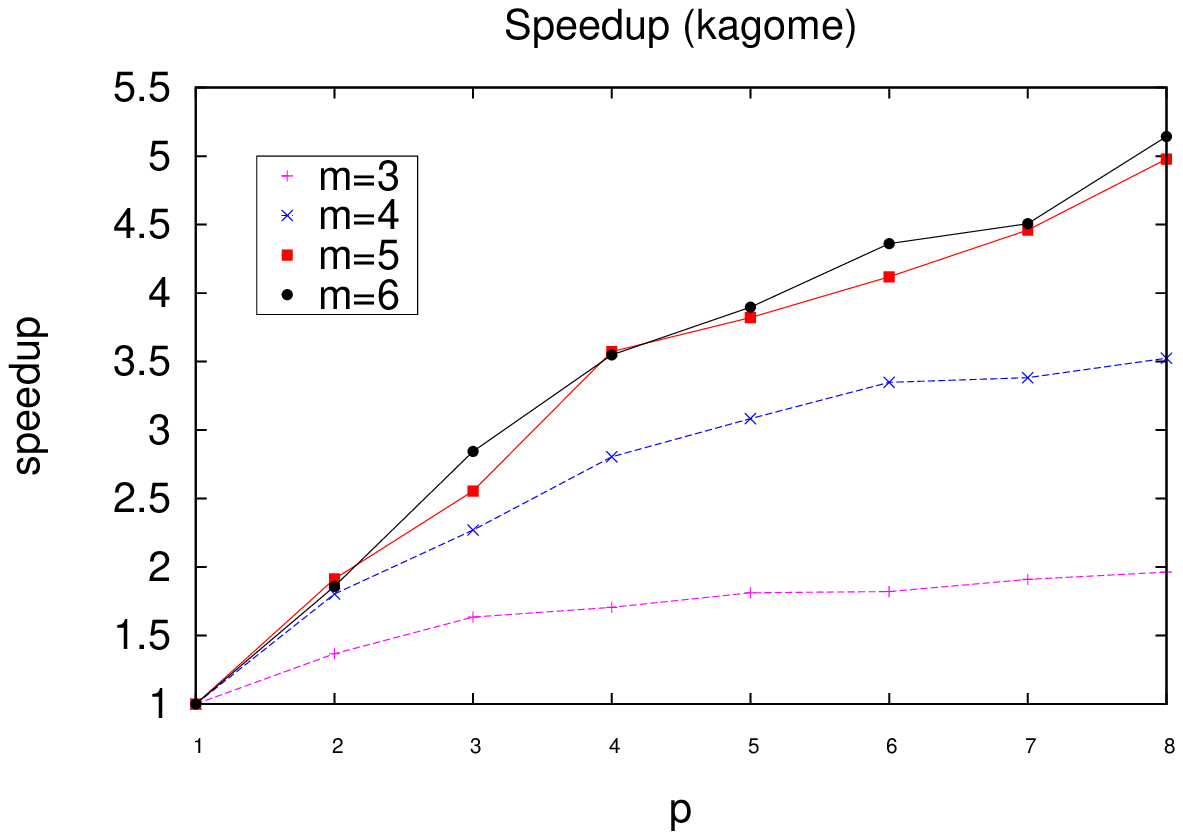}\\
\includegraphics[scale=0.7]{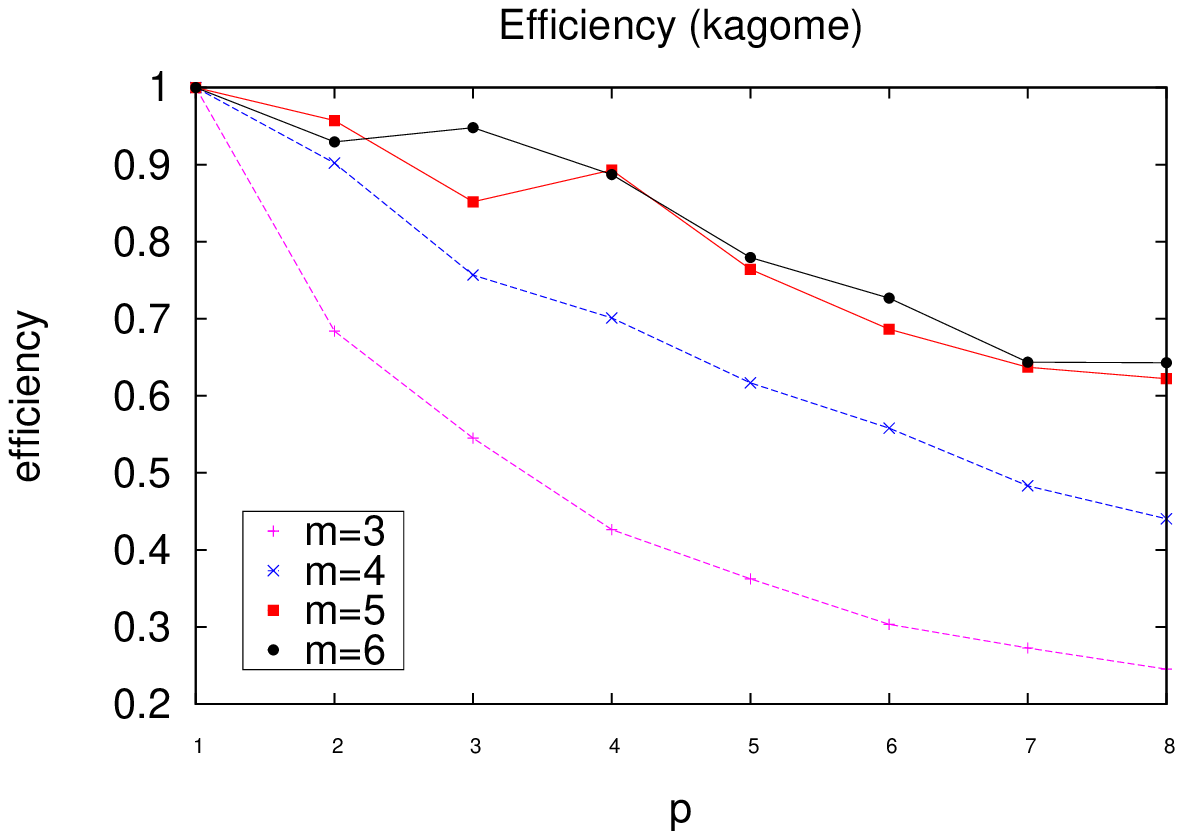} &
\includegraphics[scale=0.7]{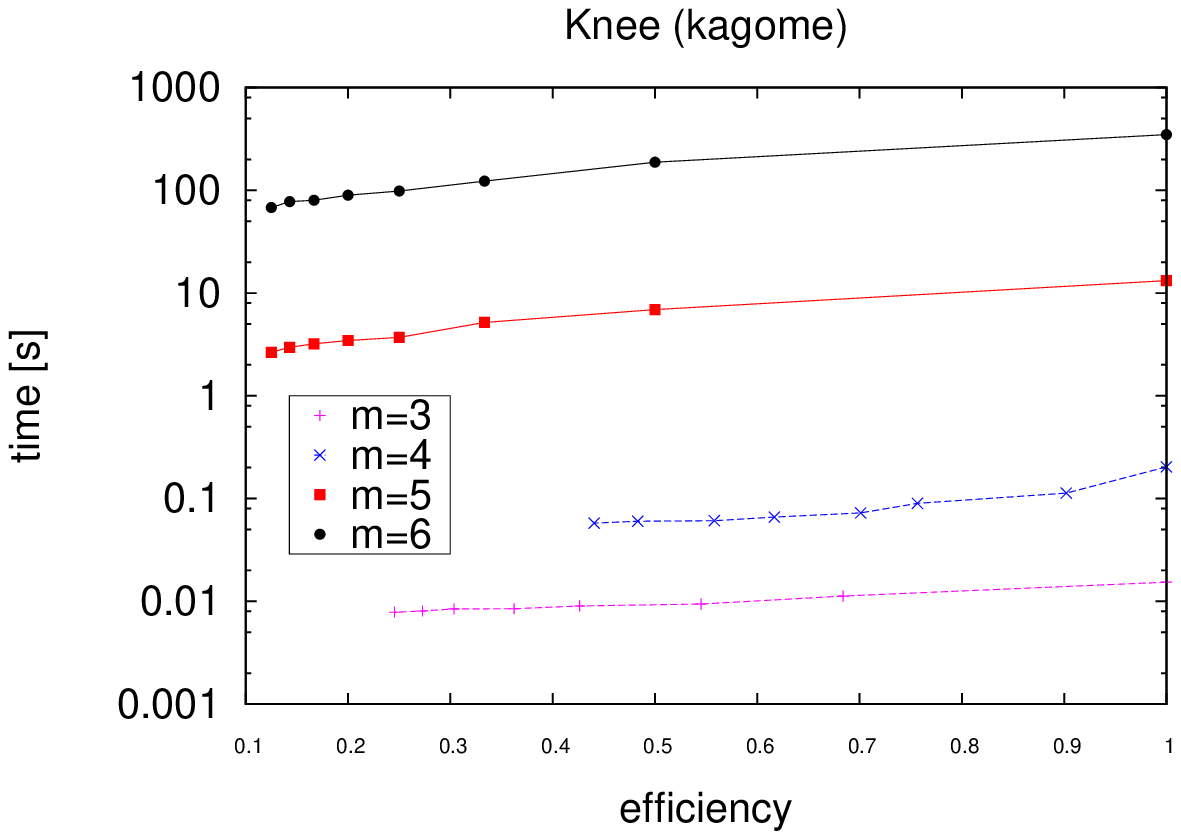}\\
\end{tabular}
\caption{Runtime, speedup, efficiency and knee for different sizes of the kagome strip lattice.}
\label{fig_kagome_performance_results}
\end{figure*}

Figure \ref{fig_kagome_performance_results} shows the performance results for the kagome strip test.
We observe that the performance of the kagome test is similar to that of the square test, but just a little lower because the deletion-contraction repetitions on the graph are not as balanced in computational cost as in the square test. Nevertheless, performance is still significantly beneficial and the maximum speedup is still $5.1$X when $p=8$ on the largest problems. When $m>5$, the efficiency of the parallel implementation is over 60\% for all values of $p$. In this test, the knee is harder to identify, but by looking into detail on the largest problems one can see a small curve that suggests $p=4$ which is in fact 90\% efficient when solving large problems. For the case of $m=4$, the knee suggests $p=2$ processors which is also 90\% efficient. 

When using symmetry on both tests, we observed an extra improvement in performance of up to $2X$ for the largest values 
of $m$. This improvement applies to both sequential and parallel execution. The size of the 
matrix transfer matrix is also improved under symmetry, in the best cases we achieved almost half the dimension of the original matrix, which in practice traduces to $1/4$ the space of the original non-symmetric matrix. Lattices as the kagome will only have certain values of $m$ where it is symmetric. In the other cases, there is no other option but to do non-symmetric computation. For lattices such as the square lattice, symmetry is always present.

\subsection{Static vs dynamic scheduler}
The implementation used for all tests used a static scheduler, hence the block-size $B$. We also implemented an alternative 
version used a dynamic scheduler. The dynamic scheduler was implemented by using a master process that handled the jobs to the worker processes. For all of our tests, the dynamic scheduler performed slower than the static scheduler. For extreme unbalanced problems, we think that the dynamic scheduler will play a more important role. For the moment, static scheduling is the best option as long as problems stay within a moderate range of work balance.

\section{Conclusions}
\label{seq_conclusions}
In this work we presented a parallel implementation for computing the transfer matrix of strip lattices in the Potts model. The implementation benefits from multi-core parallelism 
achieving up to $5.7$X of speedup with $p=8$ processors. Our most important result is the efficiency obtained for all speedup values, being the most remarkable one the $3.7$X speedup with 95\% of efficiency when using $p=4$ processors. In the presence of symmetric strip lattices, the implementation achieved an extra $2X$ of performance and used almost a quarter of the space used in a non-symmetric computation.

Our experimental results serve as an empirical proof that multi-core implementations indeed help the computation of such a complex problem as the transfer matrix for the Potts model. It was important to confirm such results not only for the classic square lattice, but also 
for more complex lattices with a higher amount of edges, such as the kagome lattice. In the kagome tests, the results 
were very similar to the ones of the square, but slightly lower because the problem becomes more unbalanced. A natural extrapolation of this behavior would suggest that very complex lattices will be even more unbalanced. We propose to use dynamic scheduling for such complex cases and static scheduling for simpler ones.

The main difficulty of this work was not the parallelization itself, but to make the problem become highly parallelizable, which is not the same. For this, we introduced a preprocessing step that generates all possible \textit{terminal configurations}, which are critical for building the matrix. This step takes 
an insignificant amount of time compared to the whole problem, making it useful in practice. 
We also introduced smaller algorithmic improvements 
to the implementation; (1) fast computation of serial and parallel paths, (2) the exploit of lattice symmetry for matrix size reduction, (3) a set of algebra rules for making consistent keys in all leaf nodes and (4) a hash table for accessing column values of the transfer matrix. 

In order to achieve a scalable parallel implementation, some small data structures were replicated among processors while some other data structures per processor were created 
within the corresponding worker process context, not in any master process. This allocation strategy results in faster cache performance and brings up the possibility for exploiting NUMA architectures.

Even when this work was aimed at multi-core architectures, we are aware that a distributed environment can become very useful to achieve even 
higher parallelism. Since rows are fully independent, sets of rows can be computed on different nodes. In the case of static scheduling (\textit{i.e.}, no master process), 
there will be no communication overhead because all processes will know their corresponding work based on their rank and the block value $B$. In the case of dynamic scheduling (\textit{i.e.}, master process), nodes will communicate sending single byte messages, and not matrix data, resulting in a small overhead which should not become a problem if the block value $B$ is well chosen. 

It is not a problem to store the matrix fragmented into many files as long as the matrix is in its symbolic form. Practical case shows that it is first necessary to evaluate the matrix on $q$ and $v$ before doing any further computation. The full matrix is needed only after numerical evaluation has been performed on every row. Again, this evaluation can be done in parallel. Under this scenario, it is evident that parallel computation of transfer matrices is highly recommendable and useful in practice. The authors of this work have achieved new exact results on a wider kagome strip lattice with the help of this implementation \cite{Alvarez_Canfora_Reyes_Riquelme_2012}.

Modern multi-core architectures have proven to be useful for improving the performance of hard problems such as the computation of the transfer matrix in the Potts model. In the future, we are interested in further improving the algorithm in order to build more efficient transfer matrices.

\section*{Acknowledgment}
The authors would like to thank \textit{CONICYT} for funding the PhD program of Crist\'obal A. Navarro. 
This work was partially supported by the FONDECYT projects $N^o$ 1120495 and $N^o$ 1120352. 

\bibliographystyle{IEEEtran}
\bibliography{tmalg_hpcc2013}
\end{document}